\documentstyle[12pt,a41,epsfig,wrapfig,rotating]{article}

\begin{document}

\begin{flushright}
hep-ph/9907210\\
DESY 99-110\\
\end{flushright}
\vspace{1.5cm}

~~~~~\\

\vspace{1cm}

\begin{center}
{\Large \bf Color Octet Contribution to $J/\psi$ Photoproduction Asymmetries }

\vspace*{2cm}
{\large G.~Japaridze}\\
\vspace*{0.5cm}
{\it 
Center for Theoretical Studies of Physical Systems, \\ 
Clark Atlanta University, Atlanta, GA 30314, U.S.}\\
\vspace*{0.8cm}
% and \\
%\vspace*{0.5cm}

\renewcommand{\thefootnote}{\fnsymbol{footnote}}
\setcounter{footnote}{1}
{\large  W.-D. Nowak and  A. Tkabladze}\footnote{Alexander von Humboldt
    Fellow} \\
\vspace*{0.5cm}
{\it DESY Zeuthen, D-15738, Zeuthen,  Germany}

\setcounter{footnote}{0}

\vspace*{3.5cm}

\end{center}

\begin{abstract}
{\small    
We investigate $J/\psi$ photoproduction asymmetries in the framework of the 
NRQCD factorization approach. It is shown that the color octet contribution
leads to  large uncertainties in the predicted asymmetries 
which rules out the possibility
to precisely measure  the gluon polarization in the nucleon 
through this final state.
For small values of the color octet parameters being  compatible with
$J/\psi$ photoproduction data it appears possible that a  measurement
 of $J/\psi$ asymmetries could provide a new test for the
 NRQCD factorization approach, on one hand, or a measurement of the polarized
gluon distribution from  low inelasticity events $(z<0.7)$, on the other.}
\end{abstract}
\vspace{1cm}

\newpage

\section{Introduction}

The study of double spin asymmetries in appropriate  final states is
a very 
promising way to measure the polarized gluon distribution function in 
the nucleon. The  favored processes are those 
for which the involved production cross sections and subprocess asymmetries 
can be calculated in the framework of perturbative QCD
and be predicted with small uncertainties.
Recently, the measurement of $J/\psi$ photoproduction asymmetries
was suggested as a tool for direct access to the  polarized gluon density 
\cite{HERMES,COMPASS}; the  analysis of asymmetries at the 
 subprocess level  was based on results obtained in the framework
 of the color singlet model (CSM) \cite{CSM,Guillet}.
However, the CSM can not  be considered as a  sufficiently
reliable model to describe heavy quarkonium production and decay processes.
The measured cross section of prompt $J/\psi$ and $\psi'$ production at 
large $p_T$ at the Tevatron exceeds the CSM predictions by  more than one order magnitude \cite{Tevatron}.
In photon-proton collisions, even after taking into account 
next-to-leading order corrections \cite{Kraemer}, the CSM does not
describe satisfactorily different distributions in $J/\psi$ 
production \cite{H1}.

The factorization approach (FA) based on 
non-relativistic QCD (NRQCD) represents a more  reliable framework to 
study heavy quarkonium production and decay processes \cite{NRQCD}. 
In an extension of the  CSM \cite{CSM}, the NRQCD FA  
implies that  the heavy quarkonium can be produced through color octet
quark-antiquark  states, as well. The production of a heavy quark-antiquark
 pair can be 
calculated perturbatively because the hard scale is 
determined by the mass of heavy quark, 
whereas the evolution of the $(Q\bar Q)$ pair to the final hadronic state
is parameterized by nonperturbative  matrix elements,
$\langle{\cal O}^{\psi}_n\rangle$.
The relative importance of the different 
intermediate states, both  color octet and color singlet,  
is   defined by  velocity scaling rules \cite{LMNMH}. 
The color singlet long distance parameters,
being related to the quarkonium wave function at the origin, can be calculated 
in the framework of  nonrelativistic models,
 while the color octet parameters have to be extracted  
from  experimental data.
%Therefore, in the FA the heavy quarkonium production
%processes is considered as double series in the QCD
%coupling constant and velocity expansion.
%Under such an assumption it is possible to get rid of infrared  
%divergences in the decays of 
%$P$-wave states into light hadrons taking account annihilation 
%of $Q\bar Q$ pair into color  
%octet $S$-state. The infrared divergences is absorbed into the color octet 
%matrix element corresponding to quarkonium transition into octet
%quark-antiquark  state \cite{BBL}. 
Once this is done,  predictions for other processes can be made.  
 Unfortunately, the present-day theoretical 
uncertainties are still too large, preventing a determination of  these
 parameters with  enough accuracy
 to check the universality of the NRQCD FA  \cite{BK}-\cite{KK}. 
%To mention briefly the sources of the 
%theoretical uncertainties note that the heavy quark pair production
% cross section
%at the Tevatron depends strongly on the factorization and/or 
%renormalization scale, on the mass of the heavy quark.

%The heavy quark pair production cross section at the
%Tevatron depends strongly on the factorization and/or renormalization scale, 
%on the mass of the heavy quark and etc. \cite{BK,BR,BRW}.

In the framework of the color octet model the main contribution to $J/\psi$
photoproduction is coming from the color octet states 
$^1S_0^{(8)}$ and $^3P_J^{(8)}$ \cite{CK,KLS}. Throughout this paper we 
exploit the fact that the corresponding long distance matrix elements can 
not be determined  separately from the 
prompt $J/\psi$ production data at the Tevatron;
it is only possible to extract their combination.

%It is obvious that in this situation  to 
%predict reliably the spin asymmetries of $J/\psi$ production and use
%this predictions to extract polarized gluon distribution 
%is not possible.

Spin effects in heavy quarkonium production, such as
$J/\psi$ \cite{BK,BKV} and $\Upsilon$ polarization \cite{KLST}-\cite{T}
in unpolarized experiments and spin asymmetries in hadroproduction
\cite{TT,GM} can provide a possibility to check the NRQCD factorization
approach. As ratios of  cross sections, spin-dependent
 observables do not strongly depend neither on the mass of heavy quarks,
nor on the  factorization and renormalization scale.
These parameters are crucial in the prediction of cross sections 
and they usually give rise to large uncertainties in the extraction of 
color octet parameters \cite{BK}.

In the present paper we consider  
double spin asymmetries in $J/\psi$ photoproduction in the framework of the 
NRQCD FA. We investigate the uncertainties depending on the 
long distance color octet parameters and the possibility to use $J/\psi$
production asymmetries for a determination of the  polarized gluon density.
We will argue that a measurement of the polarized gluon density via double spin asymmetries in 
charmonium photoproduction can potentially provide an additional test of 
the NRQCD FA, or alternatively, the polarized gluon distribution can be 
measured once the FA is confirmed. 

The deviation from the CSM prediction can also 
be used as a test of an  alternative model for heavy quarkonium production 
proposed by
Hoyer and Peigne \cite{HP} which reduces to the CSM in the case of 
$J/\psi$ photoproduction.

In the next section we consider $J/\psi$ photoproduction and electroproduction
asymmetries at  the subprocess level.
 The uncertainties in the color octet long 
distance parameters are discussed in Sec. 3. The
double spin asymmetries for various sets of long distance matrix elements
and different parameterizations of polarized parton densities are calculated 
at   $\sqrt{s}=10$ GeV which can be considered as representative for 
both the HERMES and the COMPASS experiment \cite{HERMES,COMPASS}.
We also estimate the expected double spin asymmetries in $J/\psi$
electroproduction for a possible polarized electron-proton collider with 
energy $\sqrt{s}=25$ GeV \cite{EPIC}.

\section{Double Spin Asymmetries for $J/\psi$  Subprocesses}
%at subpin $J/\psi$ photoproduction and 
%electroproduction}

\subsection{$J/\psi$ Photoproduction}
The double-spin asymmetry $A_{LL}$ for  inclusive $J/\psi$ photoproduction
 is defined as
\begin{eqnarray}    
A_{LL}^{J/\psi}(\gamma p) = \frac{
 d\sigma(\overrightarrow{\gamma}+\overrightarrow{p}\to J/\psi+X)
-d\sigma(\overrightarrow{\gamma}+\overleftarrow{p}\to J/\psi+X)}
{d\sigma(\overrightarrow{\gamma}+\overrightarrow{p}+\to J/\psi+X)
+d\sigma(\overrightarrow{\gamma}+\overleftarrow{p}\to J/\psi+X)}=
\frac{d\Delta\sigma/d^3p}{d\sigma/d^3p}~,
\end{eqnarray}
where the arrows over $\gamma$ and $p$ denote the helicity projection on
the direction of the corresponding momenta of the initial particles.

Throughout this paper we use the standard notation of spectroscopy
 to define the  quantum numbers of intermediate quark-antiquark states,
 $^{(2S+1)}L_J^{(1,8)}$. Here $S$ denotes the spin
of quark-antiquark state, $J$ is the total angular momentum, the superscripts 
$(1)$ and $(8)$ correspond to color singlet and color octet states 
respectively.
The  long distance parameter
$\langle{\cal O}_{1,8}^{H}(^{(2S+1)}L_J)\rangle$ describes the transition 
from the  $^{(2S+1)}L_J^{(1,8)}$-state to a hadron $H$.

In  lowest order of $\alpha_s$ only  color octet states contribute to
$J/\psi$ production, namely through the following subprocesses:
\begin{eqnarray} 
\gamma+g &\to& ^1S_0^{(8)}~,\\
\gamma+g &\to& ^3P_{0,2}^{(8)}
\end{eqnarray}
These heavy quarkonium states  are produced at the kinematical
endpoint in $2\to1$ subprocesses,  i.e. at  $z\simeq1$, where 
$z= P_{\psi}\cdot P_{p}/P_{\gamma}P_{p}$. 
The double-spin asymmetries for these subprocesses were calculated
in Ref. \cite{GM}. Uncontrolled uncertainties at $z\simeq1$ arise 
from higher order terms in the NRQCD velocity expansion;
in this particular 
case the higher order terms are enhanced by the kinematical factor $1/(1-z)$
which makes all predictions unreliable close to the kinematical boundary
 \cite{BRW,FM}.
In the same kinematics, the $J/\psi$ is produced via the diffractive process,
 which is an additional source of uncertainty.
Another kinematical characteristic of the $2\to1$ subprocesses, Eqs. (2)-(3),
is the small transverse momentum of the produced $J/\psi$; the transverse 
momentum of the final hadronic state is determined mainly by internal 
motion of the colliding partons and by the 
recoil of emitted soft gluons in the hadronization phase.

All these uncertainties are smaller in $J/\psi$ production at larger transverse
momenta. The leading order contribution in this case 
 comes from  $2\to2$ subprocesses which are of the order of
 $\alpha\alpha_s^2$:
\begin{eqnarray}
\gamma+g &\to& J/\psi +g~,\\ 
\gamma+g(q) &\to& ^1S_0^{(8)}+g(q),\\
\gamma+g(q) &\to& ^3S_1^{(8)}+g(q),\\
\gamma+g(q) &\to& ^3P_J^{(8)}+g(q).
\end{eqnarray}
The double spin asymmetries for  $J/\psi$ production in the 
framework of the CSM, whose corresponding subprocess is described by Eq. (4),
  were studied in Ref. \cite{Guillet}. 
We calculated the cross sections for the production of a color octet
heavy quark-antiquark pair, Eqs. (5)-(7),  
for different helicity states of initial photon and gluon (quark). 
he expressions for $\Delta\sigma$ for photon-gluon subprocesses which 
give the dominant contribution to the $J/\psi$ production
are listed in the Appendix\footnote{ The FORTRAN
codes can be provided on request}.
Our results after averaging the spins of the initial particles  are in 
agreement with those  used  for the calculation of the $J/\psi$ photoproduction
cross section in the NRQCD FA \cite{CK}\footnote{
%After we pointed out about possible mistake, Jungil Lee sent us
%recalculated results which are in agreement with [9] and our
%calculations.
 We are grateful to Michael Kr\"amer for providing us with the
analytical results for subprocess cross sections.}.
%Similar  calculations were also performed in Ref. [17].

%
\begin{figure}[th]
\vspace{-4mm}
\centering
\begin{minipage}[c]{7.5cm}
\centering
\epsfig{file=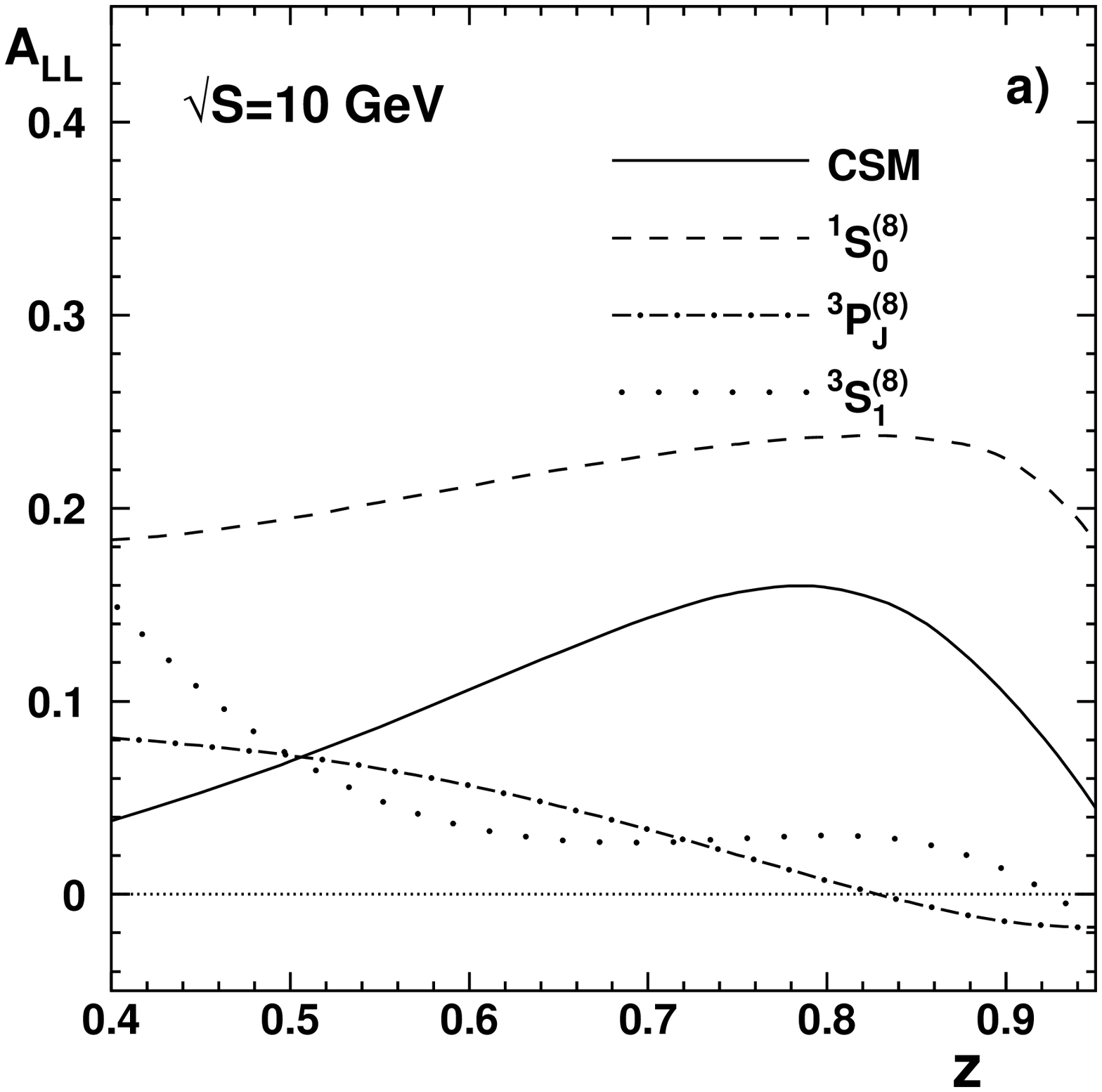,width=7.5cm}
\end{minipage}
\hspace*{0.5cm}
\begin{minipage}[c]{7.5cm}
\centering
\epsfig{file=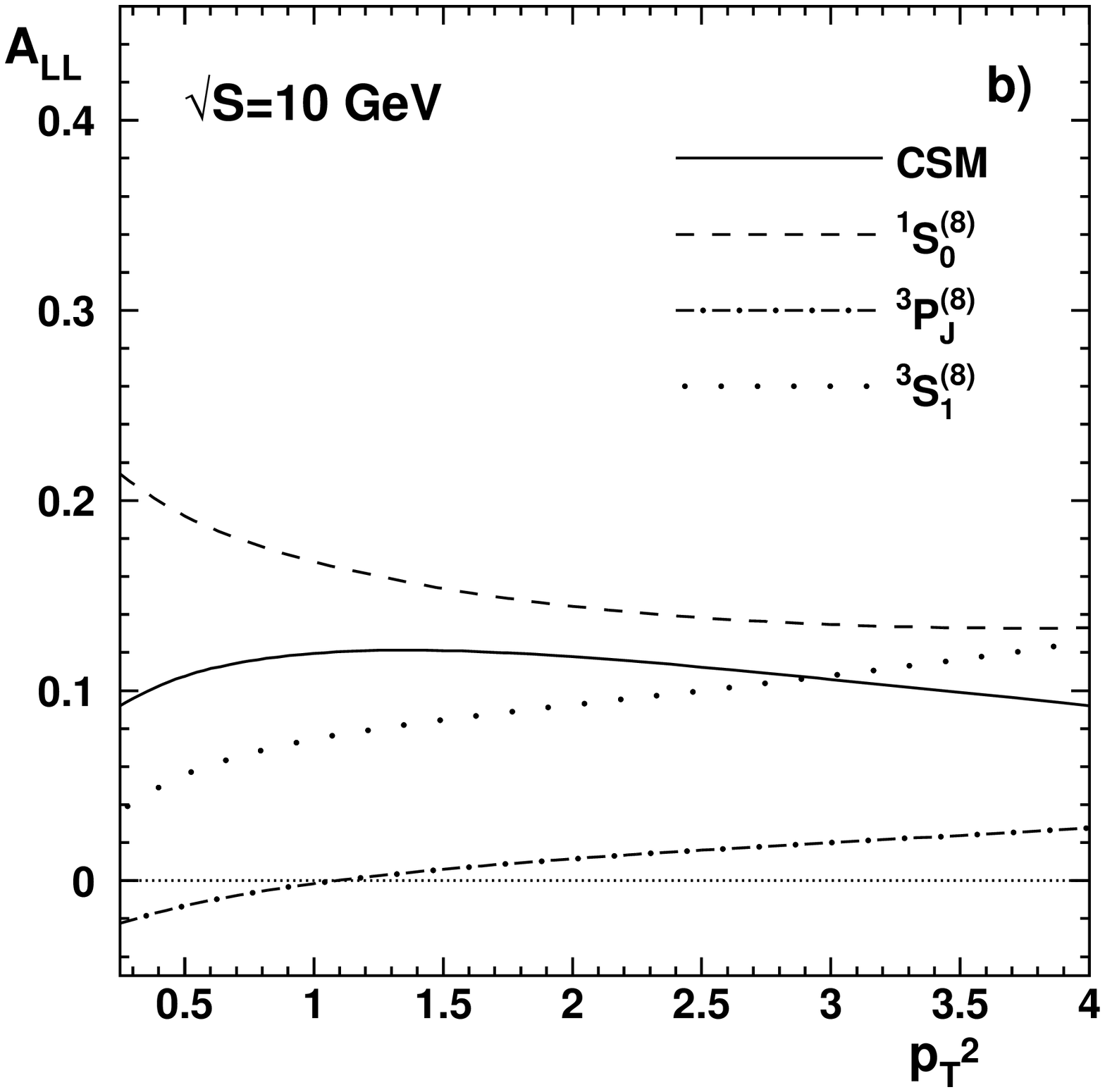,width=7.5cm}
\end{minipage}
\caption{\small
Different contributions to double spin asymmetries in quarkonium 
photoproduction,
 a) vs. $z$  and  b) vs. $p_T$, all for the GS LO 
 parameterization (set A). The solid line  
corresponds to the color singlet $^3S_1$ state,  the dashed line to the 
color octet
$^1S_0^{(8)}$ state, the dash-dotted line to the color octet $^3P_J^{(8)}$
states and the dotted line to the color octet $^3S_1^{(8)}$ state.
}
\end{figure}

The resulting double spin asymmetries for the photoproduction of various 
heavy quark-antiquark  states at $\sqrt{s}=10$ GeV
are shown in Figs.  1(a) and 1(b). They  
are  calculated for the Gehrmann-Stirling (GS) leading order (LO) 
parameterization (set A) of the
polarized parton distribution functions (PDFs) \cite{GS}.
For the unpolarized parton distribution functions the Gl\"uck-Reya-Vogt (GRV)
 LO parameterization
evaluated at $Q^2=p_T^2+4m_c^2$ was used \cite{GRV}.
As can be seen, the asymmetries for the $^1S_0^{(8)}$ (dashed lines) and 
$^3P_J^{(8)}$ (dash-dotted lines)
octet states are 
 rather different from the CSM prediction (solid line). 
Therefore, the predicted asymmetry for $J/\psi$ production  in the NRQCD FA
is expected to be sensitive to long distance color octet parameters.

\subsection{$J/\psi$ Electroproduction}

We also calculated the double spin asymmetries in heavy quarkonium 
electroproduction. 
The virtuality of the colliding photon can be used as an additional large scale in 
the process.
Requiring  $Q^2>>4 m_c^2$,  some theoretical uncertainties connected with
the enhancement of higher order corrections in the NRQCD 
expansion can be avoided.
\cite{BRW,FM}. At the same time it is possible to reduce the contribution
from higher twist effects and from $J/\psi$ diffractive production ( for a 
detailed discussion see Ref. \cite{FM}).

In the  framework of the CSM, $J/\psi$ can be produced 
  in electron-proton collisions 
 only in order $\alpha^2\alpha_s^2$ 
subprocesses, $e+g\to e+(c\bar c)+g$. The asymmetries for these subprocesses 
were calculated in Ref. \cite{Guillet}. 
However,
the main contribution to the total $J/\psi$ electroproduction cross section
comes not from CSM, but from the leading order  
$\alpha^2\alpha_s$ subprocesses, 
similar to (2),(3), when 
color octet states can be produced only with small transverse momenta
in the center-of-mass system of virtual photon and proton.
In this paper we consider the double spin asymmetries only for these
leading order subprocesses. 
The next-to-leading order color octet part, $O(\alpha^2\alpha_s^2)$,
is even smaller than the color singlet model prediction \cite{FM} and does
 not significantly change
the asymmetry calculated in the framework of the CSM in Ref. \cite{Guillet}.

In the NRQCD factorization approach in  leading order  QCD
$J/\psi$ can be produced through  two color octet states,
 $^1S_0$ and $^3P_J$.
The expression for the corresponding cross section was obtained  
in Ref. \cite{FM}:
%-----------------------------------------------------------------------
\begin{eqnarray}
&&\sigma (l+p\to l+J/\psi+X) = 
\int{\frac{ dQ^2}{Q^2}} \int{d y \frac{G(x)}{y S} } 
\; \frac{2\pi^2 \alpha_s(\mu^2) \alpha^2 e_c^2}{m_c(Q^2+4m_c^2)}
~~~~~~~~~~~~~~\nonumber \\
&&~~~~~~~~~~~~\times \Biggl\{\frac{1+(1-y)^2}{y}
 \times  \left[\langle {\cal O}_8^{J/\psi}(^1S_0)\rangle +
\frac{3Q^2+28 m_c^2}{Q^2+4m_c^2}
\frac{\langle{\cal O}_8^{J/\psi}(^3P_0)\rangle}{m_c^2}\right] \\
&&~~~~~~~~~~~~ -y\frac{32 m_c^2 Q^2}{(Q^2+4 m_c^2)^2}
\frac{\langle{\cal O}_8^{J/\psi}(^3P_0)\rangle}{m_c^2}\Biggr\},
\nonumber
\end{eqnarray}
%-----------------------------------------------------------------------
where $G(x)$ is the gluon distribution in the proton, $S$ is the 
invariant mass  of the electron-proton system, and the momentum fraction
of the struck 
parton is given by $x=(Q^2+4 m_c^2)/y S$. 

The corresponding polarized cross section has the form
%-----------------------------------------------------------------------
\begin{eqnarray}
&&\Delta\sigma(l + p\to l +J/\psi+X) = 
\frac {1}{2}(\sigma(\overrightarrow{l}\overrightarrow{p})
-\sigma(\overrightarrow{l}\overleftarrow{p})) = \nonumber \\
&&~~~~~~~~~~~~~~~\int{\frac{dQ^2}{Q^2} } \int{d y \frac{\Delta G(x)}{y S} }  
\; \frac{2\pi^2 \alpha_s(\mu^2) \alpha^2 e_c^2}{m_c(Q^2+4m_c^2)}\\
&&\times \frac{1-(1-y)^2}{y} 
 \left[\langle {\cal O}_8^{J/\psi}(^1S_0)\rangle +
\frac{3 Q^4+8m_c^2 Q^2-16 m_c^4}{(Q^2+4m_c^2)^2} 
\frac{\langle{\cal O}_8^{J/\psi}(^3P_0)\rangle}{m_c^2}\right],
\nonumber
\end{eqnarray}
%-----------------------------------------------------------------------
where $\Delta G(x)$ is the polarized gluon distribution  in the 
proton. The double spin asymmetry is defined by the ratio 
$A_{LL} = \Delta\sigma(lp)/\sigma(lp)$.

In the limit  $Q^2\to0$ the expressions for the cross sections 
reduce to the convoluted photoproduction cross sections \cite{CK,GM}
with polarized and unpolarized splitting functions correspondingly:
%-----------------------------------------------------------------------
\begin{eqnarray}
\lim_{Q^2\to0} \sigma(l+p\to J/\psi+X) &\to& \frac{\alpha}{2\pi}
\int{\frac{dQ^2}{Q^2}}\int{dy \frac{1+(1-y)^2}{y}\hat\sigma(\gamma+P\to J/\psi+X)},
\\
\lim_{Q^2\to0}\Delta\sigma(l+p\to J/\psi+X) &\to& \frac{\alpha}{2\pi}
\int{\frac{dQ^2}{Q^2}}\int{dy \frac{1-(1-y)^2}{y}\Delta\hat\sigma(\gamma+P\to J/\psi+X)}.
\end{eqnarray}
%-----------------------------------------------------------------------

\section{Matrix Elements and Results}

In $J/\psi$ photoproduction the dominant contribution comes from the
color singlet state $^3S_1$ and the color octet states
 $^1S_0^{(8)}$ and $^3P_J^{(8)}$.
The production cross section 
for the $^3S_1^{(8)}$ octet state is much smaller than the cross sections 
for these two octet states \cite{CK}.
 The color singlet matrix element is connected to the nonrelativistic wave
 function at the origin and can be evaluated in the framework 
of potential models. It can also be extracted from the leptonic decay width
 of the $J/\psi$. 
The corresponding color octet long distance matrix elements
 can not be calculated 
in  perturbative QCD (pQCD) and are usually extracted from  experimental data.
For the color octet $P$-wave states they
can be reduced to one parameter using the NRQCD spin symmetry relation:
%-----------------------------------------------------------------------
\begin{eqnarray}
\langle{\cal O}_8^H(^3P_J)\rangle & = & (2J+1) 
\langle{\cal O}_8^H(^3P_0)\rangle.
\end{eqnarray}
%-----------------------------------------------------------------------

After using these relations there are only two matrix elements left 
which give the main octet contribution to $J/\psi$ photoproduction,
$\langle{\cal O}_8^{\psi}(^1S_0)\rangle$ and 
$\langle{\cal O}_8^{\psi}(^3P_0)\rangle$.
From the available  data of prompt $J/\psi$ production at the Tevatron
it is only possible to extract  the combination
of these matrix elements with rather large uncertainties \cite{CL,BK,KK}.
Apparently, at Tevatron energies  the states
 $^1S_0^{(8)}$ and $^3P_J^{(8)}$ contribute
significantly 
at small transverse momenta of $J/\psi$, $p_T<5$ GeV.
Consequently, the fitted value of the combination of matrix elements 
$\langle{\cal O}_8^{\psi}(^1S_0)\rangle$ and 
$\langle{\cal O}_8^{\psi}(^3P_0)\rangle$, defined as \cite{KK}
%------------------------------------------------------------------------- 
\begin{eqnarray}
M_r^{J/\psi} &=& \langle{\cal O}_8^{J/\psi}(^1S_0)\rangle
+\frac{r\cdot\ \langle{\cal O}_8^{J/\psi}(^3P_0)\rangle}{m_c^2}~,
\end{eqnarray}
%------------------------------------------------------------------------- 
is very sensitive to effects 
which change the $p_T$ spectrum, such as dependence on factorization and/or
renormalization scales, choice of parameterization of parton distribution 
 functions 
etc. \cite{BK}. The number $r$ is fitted from experimental data.

To test the sensitivity of double spin asymmetries on the color octet matrix
elements we used  two sets of parameters extracted from the
Tevatron data on prompt $J/\psi$ hadroproduction \cite{KK}, as shown in
 Table I:
%------------------------------------------------------------------------- 
\begin{center} 
\begin{tabular}{c|c|c|c} 
\hline \hline 
          & LO   &  HO   & scaling \\ 
\hline 
$\langle{\cal O}_1^{J/\psi}(^3S_1)\rangle$ & 
$(7.63\pm0.54)\cdot10^{-1}$ GeV$^3$ & $ (1.30\pm0.09)$ GeV$^3$ &
$[m_c^3 v^3]$ \\
$\langle{\cal O}_8^{J/\psi}(^3S_1)\rangle$ & 
$(3.94\pm0.63)\cdot10^{-3}$ GeV$^3$ & $ (2.73\pm0.45)\cdot10^{-3}$ GeV$^3$ &
$[m_c^3 v^7]$ \\
$M_r^{J/\psi}$  & 
$(6.52\pm0.67)\cdot10^{-2}$ GeV$^3$ & $ (5.72\pm1.84)\cdot10^{-3}$ GeV$^3$ &
$[m_c^3 v^7]$ \\
$r$ & 3.47 & 3.54 &  \\
\hline
\end{tabular} 
\end{center} 
 
\noindent 
 Table I. {\small Values of the long distance matrix elements from LO- and
 HO-improved fits to the CDF data.}
\vspace{0.3cm} 
%------------------------------------------------------------------------- 

\noindent
The LO  set of the matrix elements corresponds to the leading-order  fit
of the prompt $J/\psi$ hadroproduction data at the Tevatron and the obtained 
value for $M_r^{J/\psi}$ is the largest among existing fits in the
literature. These values of matrix elements lead to an increase 
of the photoproduction  cross section as $z\to1$ and dramatically 
overestimate the H1 and ZEUS data on $J/\psi$  photoproduction at HERA.

\begin{figure}[ht]
\vspace{-4mm}
\centering
\begin{minipage}[c]{7.5cm}
\centering
\epsfig{file=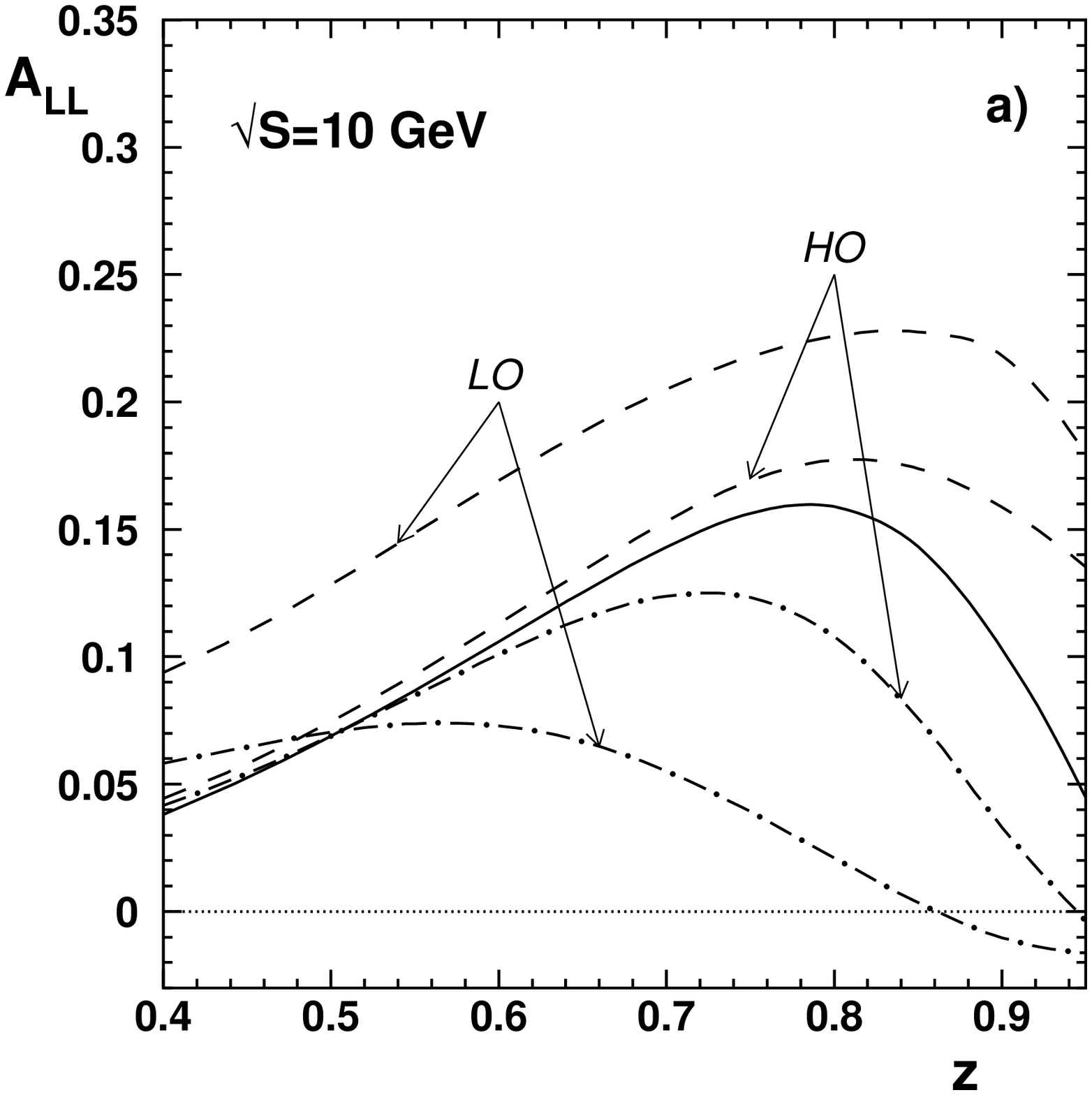,width=7.5cm}
\end{minipage}
\hspace*{0.5cm}
\begin{minipage}[c]{7.5cm}
\centering
\epsfig{file=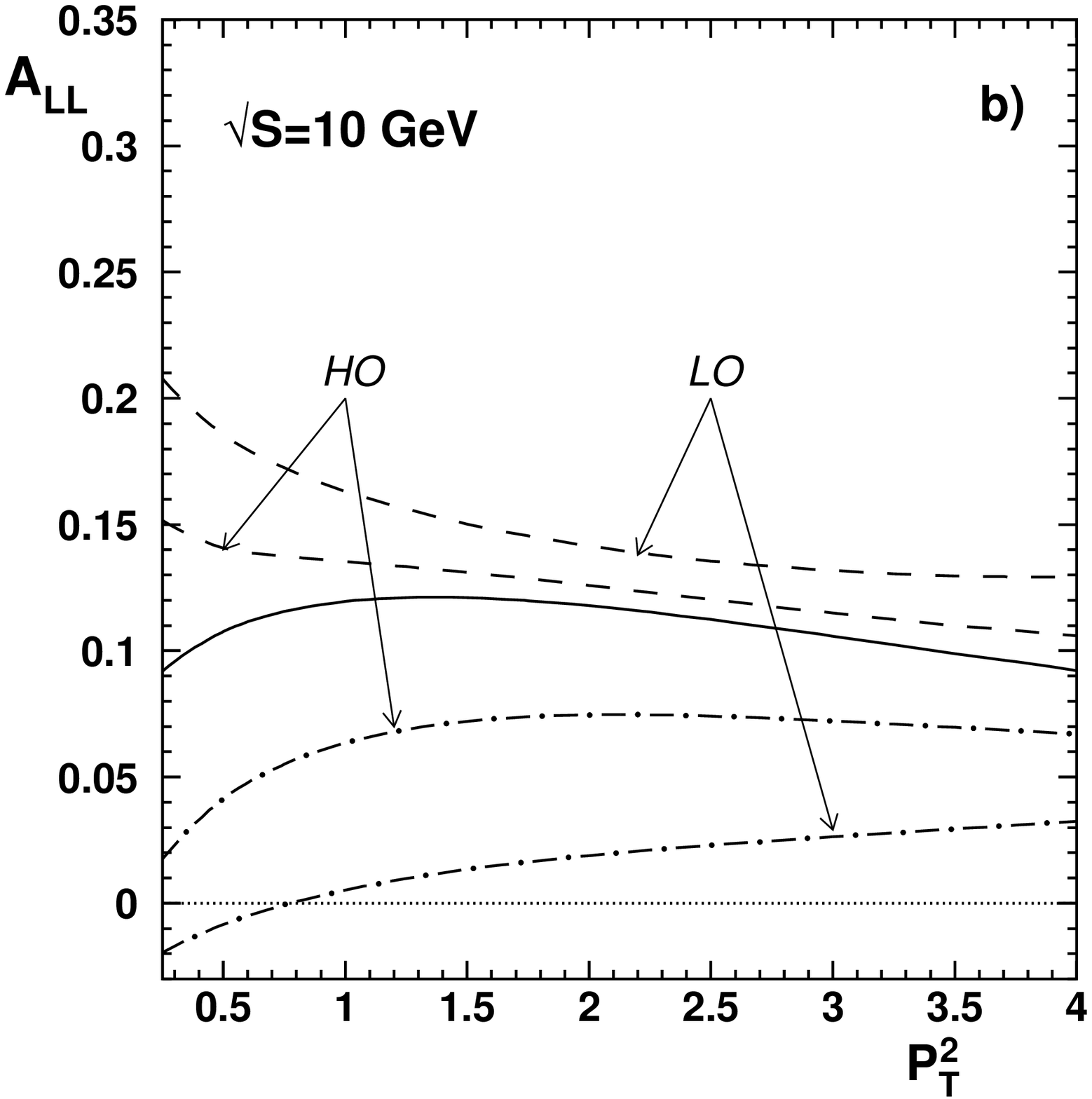,width=7.5cm}
\end{minipage}
\caption{\small
  (a) $z$ and  (b) $p_T$ dependence of double spin asymmetries of
$J/\psi$ photoproduction
for LO and HO sets of color octet matrix elements. The solid curves correspond 
to the CSM prediction. The dashed curves correspond to the case when 
$\langle{\cal O}_8^{J/\psi}(^1P_0)\rangle = 0$, the dash-dotted ones 
to  $\langle{\cal O}_8^{J/\psi}(^1S_0)\rangle = 0$.
}
\end{figure}

When  fitting the  HO set, the higher-order  corrections were taken into 
account only 
approximately by considering the initial and final state gluon radiation
in the framework of a Monte Carlo simulation \cite{CS,KK}. 
Including effectively the primordial transverse momentum of partons 
increases the production cross sections of $^1S_0^{(8)}$ and $^3P_J^{(8)}$ 
color octet states at small values of $p_T$. The impact of this effect is a
significant decrease of the $M_r^{J/\psi}$ combination by about one order of 
magnitude, as can be seen from table I.
It is worth  noting that this 
 type of HO corrections changes the photoproduction cross
sections for the color octet states only by about $20\%-30\%$ 
at  H1 and ZEUS energies \cite{KK}.
The HO parameters, presented in Table I, are the smallest ones from the 
existing fits and are compatible with the H1 and ZEUS data on the 
$z$-distribution of the $J/\psi$ {\it photo}production cross section.
We consider this set as an example of small color octet parameters.
The combination of color octet parameters $M_r^{J/\psi}$ was
 also extracted from
fixed target {\it hadro}production data \cite{BR,GuS} for $r=7$,
 $M_7^{J/\psi}\simeq 2\div3 10^{-2}$ GeV$^3$. We note that these values lie in
between the LO and HO sets of parameters, as given Table I, 
We used this range  to check the sensitivity
of the expected asymmetries on the model of $J/\psi$ production.

In  Fig. 2(a) and 2(b) the expected asymmetries in $J/\psi$ photoproduction, 
calculated at  $\sqrt{s}=10$ GeV
for the   GS LO  set A parameterization of polarized PDF's,
are shown in dependence on $z$ and $p_T^2$.
%For unpolarized PDF the GRV LO parameterization was used \cite{GRV}.
As can be seen  from  these figures, the uncertainty
in the asymmetry of $J/\psi$ production is large for the LO set of color octet 
matrix elements because  the  values of 
$\langle{\cal O}_8^{J/\psi}(^1S_0)\rangle$ and 
$\langle{\cal O}_8^{J/\psi}(^3P_0)\rangle$ are not fixed separately but only
 as the combination (13).
For the HO set
the contribution to the cross section 
of the $^1S_0^{(8)}$ and $^3P_J^{(8)}$ color octet states is smaller, which 
reduces the deviation from the  asymmetry predicted by the CSM 
especially for lower $z$-values $(z<0.7)$.
To make a choice between the different
 sets of long distance
parameters,   more accurate measurements at energies lower  than that for
H1 and ZEUS  are required in the future.
The COMPASS experiment at CERN \cite{COMPASS} and later possibly a
 presently discussed polarized electron-proton  collider (EPIC) 
at  $\sqrt{s}=25$ GeV \cite{EPIC} will work at higher
luminosities and may hence  provide such  measurements.
The gluon polarization itself at that time has  probably been  measured in 
semi-inclusive reactions with less uncertainties, for example in 
open charm production at COMPASS  or in dijet and direct photon
 production at RHIC \cite{RHIC}.
If, in this situation,   the HO set of parameters should  fit the  data of 
the lower energy experiments, the
$J/\psi$ photoproduction asymmetries can be used to test the NRQCD FA.
Alternatively, based upon the HO set of parameters and relying on the NRQCD FA 
approach, the polarized gluon distribution can be measured with 
acceptable theoretical uncertainties, if only $J/\psi$ events with 
$z<0.7$ will be used.
\begin{wrapfigure}{l}{7.5cm}
\vspace*{-4mm} 
\centering
\epsfig{file=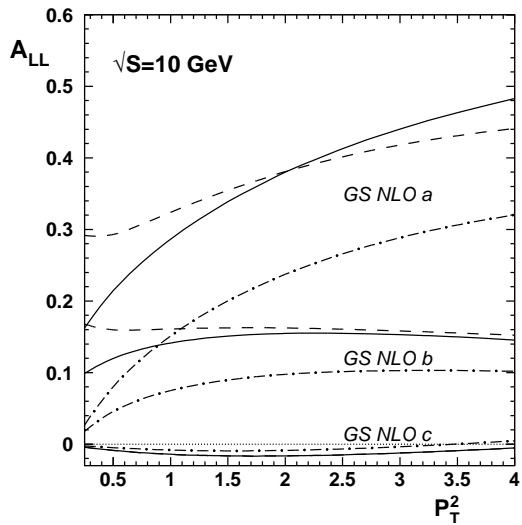,width=7.5cm}
\vspace{-7mm}
\caption{\small 
The double spin asymmetries for $J/\psi$ photoproduction at
$\sqrt{s}=10$ GeV using HO long distance parameters 
versus $p_T$.
The definition of lines is the same as in the previous figure.
}
\end{wrapfigure}

Using the HO set of color octet parameters in  Fig.~3 the double spin
 asymmetries for $J/\psi$ photoproduction
are presented  at $\sqrt{s}=10$ GeV 
for the  different sets A, B, C of 
GS NLO parameterizations for polarized PDF's \cite{GS}.
We used here different  sets of NLO parameterization to show the sensitivity of
the asymmetries to the polarized gluon distribution function, which is rather
 different for these three sets.
We note that, as expected, the    result for the GS LO  set A 
parameterization  is practically the same as for the GS NLO set B.
 We show only the $p_T^2$ dependence of the asymmetries.
%The large asymmetries for $J/\psi$ production are expected at large values of
%$z$. On the other hand,
As  was mentioned above,   uncontrolled
corrections are expected at large $z$  from higher order terms in the velocity 
expansion  because of ignoring the mass difference between the 
$(c\bar c)$-pair and the final state hadron. 
This type of uncertainties can be neglected in the $d\sigma/dp_T$
distribution because the integration over $z$ smears the singular region out
\cite{BRW,FM}. 
 When using the HO set of  color octet parameters, 
the $p_T$ behaviour of the 
 double
spin asymmetry for $J/\psi$ production 
is rather well described by the CSM prediction as long as color octet 
$P$ states don't contribute, as can be seen from Fig. 3.
 If, in contrast, the color octet $S$-state does
not contribute, the asymmetry decreases by about $30\%$. This behaviour 
appears for $p_T>1$ GeV and seems to be rather independent of the actual NLO 
parameterization set used for polarized PDF's.

In $J/\psi$ photoproduction experiments at low energy
higher twist effects are expected to be 
suppressed by the  heavy quark mass, which defines the hard scale of the 
corresponding subprocess, ~$\Lambda_{QCD}/m_c$. 
Moreover, it is not excluded that higher twist effects are suppressed even 
stronger by the transverse momentum of the produced quarkonium statee,
 or by factor $(m_{Q\bar Q}^2+p_T^2)^{-1/2}$. 
A study of these effects may be performed 
at high luminosity experiments such as eN at DESY TeV Energy Superconductiong 
Linear Accelerator TESLA,\footnote{A project of
a high luminosity experiment  scattering of  TESLA electron beam off
polarized nucleons is under consideration \cite{WD}.} where enough statistics 
can be accumulated at larger values of transverse momentum.

%are rather different even for the 
%medium size gluon polarization in the nucleon.
%The deviation from the CSM prediction for double spin
%asymmetries is not large in this case.
%
%Hence,  the measurement of $J/\psi$ production 
%asymmetries can be used to discriminate between different gluon polarizations
% in the
%nucleon if the color octet parameters should be  established and small.
%Moreover, the  measurement of double distribution cross section, 
%$d\sigma/dzdp_T$, may then even  allow to access directly  
%$\Delta G(x)/G(x)$ \cite{Guillet}, 
%with acceptable theoretical uncertainties by 
%assuming CSM dominance.

In  Fig. 4  electroproduction asymmetries are shown at 
$\sqrt{s}=25$ versus photon virtuality $Q^2$, calculated using the 'largest'
 parameterization, namely GS NLO set A. 
 As can be seen from this figure,
the difference between the asymmetries for the color octet  states
 $^1S_0^{(8)}$ and $^3P_J^{(8)}$ is large for a large gluon polarization
offering a powerful tool to discriminate between the 
hitherto 'coupled' contributions from color octet $S$- and $P$-wave states.
It is important to note  that 
in  leading order electroproduction 
large contributions from higher twist effects and diffractive $J/\psi$
production are expected. They can be  suppressed 
by requiring 
large enough $Q^2>>4 m_c^2$  which leads, however,
to  a rapidly falling  cross section  and 
large statistical errors are expected \cite{FM}.
It is a question of attainable statistics at  EPIC 
whether it may be  possible to achieve  reasonable restrictions on the values 
of the color octet parameters from  $J/\psi$ electroproduction 
asymmetries.
At  high energies, $\sqrt{s}>100$ GeV, which correspond to the polarized 
HERA option \cite{HERA}, the total cross section is large
but the asymmetries calculated with the 
commonly used parameterizations GS or Gl\"uck-Reya-Stratmann-Volgelsang
(GRSV) \cite{GS, GRSV} are expected to be very small, less than $2\%$. 
%At such  high energies the characteristic 
%values of partonic $x$ are small, $O(10^{-2}),
% yielding  where the polarized gluon density 
%is expected to be  very small, as seen from today.

\section{Conclusions}

In the present paper  double spin asymmetries in $J/\psi$
photoproduction
 and electroproduction were studied in the NRQCD factorization approach.
 The asymmetries
for the color octet states $^1S_0^{(8)}$ and $^3P_J^{(8)}$ which give
the dominant octet 
 contribution to the $J/\psi$ {\it photo}production cross section, are
significantly different from the CSM prediction.
The values of the corresponding color octet long distance matrix elements
 can presently not be extracted 
separately from  existing data on $J/\psi$ hadroproduction.
Two sets of long distance parameters were used to show the sensitivity 
\begin{wrapfigure}{r}{7.5cm}
\vspace*{-10mm}
\centering
\epsfig{file=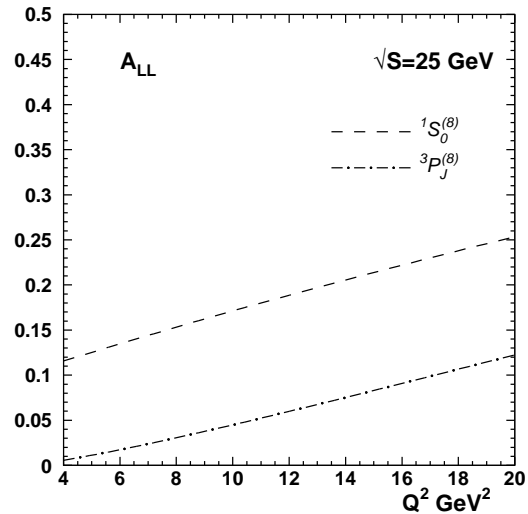,width=7.5cm}
\vspace{-7mm}
\caption{\small 
The Double spin asymmetries for $J/\psi$ electroproduction at $\sqrt{s}=25$ GeV
for GS NLO (set a) parameterization.
The dashed  line corresponds to the asymmetry of $^1S_0^{(8)}$ state and
dash-dotted line to the asymmetries of $^3P_J^{(8)}$ states.
}
\end{wrapfigure}
of the expected asymmetries to this uncertainty. 
For the 'large' set of parameters extracted from CDF data \cite{KK}
the asymmetries contain very large uncertainties. 
For the 'small'  set of 
parameters,
when higher order QCD corrections are partly  taken into account, the 
predicted asymmetries for the two extreme  choices of parameters are 
for $z<0.7$ rather close
to the CSM prediction. This 'small' set of long distance parameters is 
preferable to explain  H1 and ZEUS data on inelastic $J/\psi$ 
photoproduction. The planned high luminosity measurements at 
COMPASS   and 
possibly at EPIC ($\sqrt{s}=25$ GeV)
may provide the  possibility to fit the color octet parameters
in {\it electro}production at 
lower energies with better accuracy.
Here even a discrimination between $S$-wave and $P$-wave color octet
states appears feasible because the CSM contribution is subleading in 
electroproduction.
If the 'small' set of parameters should remain  the prefered one,
the $J/\psi$ {\it photo}production asymmetries
can be eventually used to test the NRQCD FA.
Here it is  anticipated  that the gluon polarization
will have  been  measured already with small uncertainties 
through  other semi-inclusive modes in polarized 
electron-proton or proton-proton collisions.
%In this case $J/\psi$ production can also be used to access $\Delta G/G$
%via CSM dominance with not too large theoretical uncertainties.
On the other hand, once the situation with the NRQCD FA has been clarified 
by other experiments, such as $J/\psi$ and bottomonium 
polarization measurements,
the measurement of double spin asymmetries in
$J/\psi$  {\it photo}production at HERMES and COMPASS 
can be used to extract the gluon
polarization in the nucleon in  LO  pQCD using events in the range
 $0.4<z<0.7$, because for $z<0.4$
the contribution of resolved photon subprocesses becomes essential.

\vspace{0.3cm}

{\bf Acknowledgment} We are grateful to Michael Kr\"amer for stimulating
and helpful discussions. This work is supported partly by the Alexander von
Humboldt foundation and by the National Science Foundation.

\newpage
\setcounter{equation}{0}
\def\theequation{A.\arabic{equation}}

\noindent
{\Large\bf 
Appendix}

\vspace{0.3cm}
\noindent
Here we present the polarized cross sections for  
the production of $(c\bar c)$ states in photon-gluon fusion subprocesses. 
The polarized cross section is defined as follows
\begin{eqnarray}
\Delta\sigma & =& \frac{\sigma(++)-\sigma(+-)}{2},
\end{eqnarray}
where the first and secod signs denote 
 the helicity states of colliding photon and gluon, respectively.
The expressions for unpolarized cross sections are listed in Ref. \cite{KLS}.
All cross sections were calculated for different helicity states of 
colliding particles. The obtained results were checked by comparing 
the unpolarized cross sections of this work, $(\sigma(++)+\sigma(+-))/2$,
 with those 
obtained in \cite{CK,KLS}. 
As an additional check, we calculated the asymmetries for the 
production of scalar,
 $^1S_0^{(8)}$ and $^3P_0^{(8)}$, and tensor $^3P_0^{(8)}$ 
octet states in the limit
$s\to M_{J/\psi}$
 (i.e., at the threshold of  quarkonium pair production).
The subprocess level asymmetries for scalar and tensor states 
approach 1 and -1,
respectively. The limit $4 m^2_c\to s$ implies that the emitted gluon in the 
$2\to2$ subprocess $\gamma g\to(c\bar c)g$
becomes soft; the helicity properties of the amplitude then become the same
as those of the amplitude of the $2\to1$ process $\gamma g\to(c\bar
c)$. 
%\footnote{The same reason makes the asymmetry less sensitive
% to the radiative corrections, than spin-averaged cross section, as K-factor
% is dominated by the contributions of virtual, soft and collinear gluons}. 
It is easy to show that the asymmetries for  the process $2\to1$ 
for scalar and tensor states are 1 and -1, respectively;
 the scalar(tensor) states
are produced only in the case when the helicities of initial
 photon and gluon are
$[++]([+-])$,  which means that they  have a total momentum 
projection $J_z=0(\pm2)$. 
The existence of such limits serves as an additional test for
the  analytical calculations of the polarized cross sections $\Delta\sigma$.

The spin dependent differential cross section for $J/\psi$ production  through a
color singlet state has the form
\begin{eqnarray}
\frac{\Delta\sigma(\gamma g\to J/\psi\;g)}{dt} =
\frac{(ee_cg_s)^2\langle{\cal O}_1^{J/\psi}(^3S_1)\rangle}
{16 \pi s^2} \frac{32}{27 Q^2} \big\{(st+tu+su)^2-M^2stu\bigr\},
\end{eqnarray}
with $Q=(s+t)(s+u)(t+u)$.

 For the case of an intermediate $^1S_0^{(8)}$ color octet state the
corresponding cross section reads as

\begin{eqnarray}
&&\frac{\Delta\sigma(\gamma g\to ^1S_0^{(8)} g\to J/\psi X)}{dt} =
\frac{(ee_cg_s^2)^2\langle{\cal O}_8^{J/\psi}(^1S_0)\rangle}
{16 \pi M t Q^2} 3s u(s^4-t^4-u^4+M^8).
\end{eqnarray}
For  $J/\psi$ production through the color octet state $^3S_1^{(8)}$
the polarized cross section 
can be obtained from (A.2) by changing the long distance
matrix element:
$\langle{\cal O}^{J/\psi}(^3S_1)\rangle\to
15/8\langle{\cal O}_8^{J/\psi}(^3S_1)\rangle$.

Spin dependent cross sections for  $^3P_J^{(8)}$
octet states have the form:
\begin{eqnarray}
&&\frac{\Delta\sigma(\gamma g\to ^3P_0^{(8)} g\to J/\psi X)}{dt} =
\frac{(ee_cg_s^2)^2\langle{\cal O}_8^{J/\psi}(^3P_0)\rangle}{\pi s^2Q^3 t}
384\biggl\{ 4 M^6 s1^{-1} t^5 t_1 t_2 
\nonumber\\
&&~~~ +4 M^6 t^4 t_2 (3 t-2 M^2)- M^2 s_1 t^3  (3 t^4+2 t^3 M^2-41 t^2 M^4
\nonumber\\
&&~~~ -14  t M^6+3 M^8) 
-M^2 s s_1^2 t^3  (41 t^2-118 t M^2+68 M^4) 
\nonumber\\
&&~~~ -M^2 s^2 s_1^2 t  (73 t^3-191 t^2 M^2+139 t M^4-45 M^6) 
\nonumber\\
&&~~~ -4 M^2 s^3 s_1^2 t  (19 t^2-42 t M^2+24 M^4) 
-9 M^2 s^4 s_1^2  (5 t^2-9 t M^2+3 M^4) 
\nonumber\\
&&~~~ -12 M^2 s^5 s_1^2  t - M^2 s_1^2 t^3 
 (14 t^3-33 t^2 M^2-22 t M^4+3 M^6) 
\nonumber\\
&&~~~ -s_1^2 t^2 (t^2+t s+2 s^2) u_1^3 
 +9 M^{10}  s^2 s_1 t +18 M^4 s s_1^2  ( M^4(t^2+s^2)+s^4)\biggr\}
\end{eqnarray}

\begin{eqnarray}
&&\frac{\Delta\sigma(\gamma g\to ^3P_1^{(8)} g\to J/\psi X)}{dt} =
\frac{(ee_cg_s^2)^2\langle{\cal O}_8^{J/\psi}(^3P_1)\rangle}{\pi s^2Q^3 t}
384\biggl\{ -4M^6 s_1^{-1} t^5 t_1 t_2
\nonumber\\
&&~~~ +4 M^4 t^5 (t^2-3 t M^2+6 M^4)+2 M^2 s_1 t^4 t_2 (28 M^4+5 t t_1) 
\nonumber\\
&&~~~ +2 M^2 s s_1^2 t^3 (13 t^2+20 t M^2+3 M^4)
+2 M^2 s^2 s_1^2 t^2  (7 t^2+11 t M^2-2 M^4)
\nonumber\\
&&~~~ +12 M^2 s^3 s_1^2 t^2 u+2 M^2 s_1^2 t^4 (11 t^2+16 t M^2+23 M^4)
\nonumber\\
&&~~~ -4 s_1^2 t^2 (t^2+t s+2 s^2) u_1^3
+8 M^4 t^3 (M^4 s_1^2+2 t^3 s_1-M^6 u)\biggr\}
\end{eqnarray}

\begin{eqnarray}
&&\frac{\Delta\sigma(\gamma g\to ^3P_2^{(8)} g\to J/\psi X)}{dt} =
\frac{(ee_cg_s^2)^2\langle{\cal O}_8^{J/\psi}(^3P_2)\rangle}{\pi s^2Q^3 t}
128\biggl\{ 2 M^6 s_1^{-1} t^5 t_1 t2
\nonumber\\
&&~~~ +2 M^4 t^3 (3 t^3 t2+10 t t2 M^4-12 M^6 t_1)
+  M^2  s_1 t^4 (3 t^3+20 t^2 M^2
\nonumber\\
&&~~~ +85 t M^4+52 M^6)+M^2 s s_1^2 t^2 (11 t^3+116 t^2 M^2-55 t M^4+48 M^6) 
\nonumber\\
&&~~~ +M^2 s^2 s_1^2 t^2  (t+2 M^2) (13 t+11 M^2) 
-2 M^2 s^3 s_1^2 t  (t^2+39 t M^2-48 M^4) 
\nonumber\\
&&~~~ -6 s^4 s_1^2 t M^2 (t+14 M^2) -24 s_1^2 M^4 s^5 
\nonumber\\
&&~~~ + M^2 s_1^2 t^3
 (5 t^3+72 t^2 M^2+53 t M^4+12 M^6) 
\nonumber\\
&&~~~ -2 s_1^2 t^2 (t^2+t s+2 s^2) u_1^3
+12 M^8 s^2 s_1  (2 s u+t M^2)
\nonumber\\
&&~~~ +36 M^6 s_1 (M^4 t^3+s^5-s^4 M^2+M^4 t^2 s)\biggr\}
\end{eqnarray}

\vspace{0.2cm}
The same process has been considered
 in another paper \cite{China}. We note that our 
 results for unpolarized and polarized cross sections for
$^3P_0^{(8)}$ octet state production and for the spin dependent cross section
of the $^3P_2^{(8)}$ octet state are different compared to Ref. \cite{China}.
However, our
results for the sums of cross sections for $P$ states,
both polarized and unpolarized,  were confirmed by those obtained in 
\cite{China}.

\end{document}